\documentclass{aa}  

\usepackage{graphicx}
\usepackage{txfonts}
\usepackage{natbib,twoopt}
\usepackage{amsmath}
\usepackage{multirow}
\usepackage{float}
\usepackage{lineno}
\begin{document}

   \title{Spatial distribution of isotopes and compositional mixing in the inner protoplanetary disk}

   \author{Kang Shuai
          \inst{1}
          \and
          Hejiu Hui\inst{1}\fnmsep\inst{2}\fnmsep\inst{3}
          \and
          Li-Yong Zhou\inst{4}
          \and
          Weiqiang Li\inst{1}\fnmsep\inst{2}
          }

   \institute
             {State Key Laboratory for Mineral Deposits Research \& Lunar and Planetary Science Institute, School of Earth Sciences and Engineering, Nanjing University, Nanjing,
               210023, China\\
              \email{hhui@nju.edu.cn}
         \and
             CAS Center for Excellence in Comparative Planetology, Hefei, 230026, China
         \and
             Key Laboratory of Earth and Planetary Physics, Institute of Geology and Geophysics, Chinese Academy of Sciences, Beijing, 100029, China
         \and
             School of Astronomy and Space Science, Nanjing University, Nanjing, 210023, China
             }

   \date{}

 
  \abstract
  {The mass-independent isotopic signatures of planetary bodies have been widely used to trace the mixing process that occurred during planet formation. The observed isotopic variations among meteorite parent bodies have been further linked to the modeled mass-weighted mean initial semimajor axes in N-body simulations, assuming a spatial isotopic gradient in the inner protoplanetary disk. However, nucleosynthetic isotopic anomalies of nonvolatile elements and mass-independent oxygen isotopic variation ($\Delta ^{17}$O) show different relationships with distance from the Sun. Therefore, it is crucial to know whether isotopes were distributed systematically with heliocentric distance in the inner protoplanetary disk. In this study, we performed N-body simulations on compositional mixing during the collisional accretion and migration of planetary bodies to investigate the spatial distributions of Cr and O isotopes in the inner protoplanetary disk. The modeled mass-weighted mean initial semimajor axes of the parent bodies of noncarbonaceous (NC) meteorites and terrestrial planets were used to calculate the isotopic compositions of these bodies. Our simulations successfully reproduced the observed nucleosynthetic Cr isotopic anomaly among Earth, Mars, and the NC meteorite parent bodies, consistent with a spatial gradient of isotopic anomalies in the inner disk. Asteroids originating from different regions in the inner disk were transported to the main belt in our simulations, resulting in the Cr isotopic anomaly variation of the NC meteorite parent bodies. However, the $\Delta ^{17}$O distribution among the terrestrial planets and the NC meteorite parent bodies could not be reproduced assuming a $\Delta ^{17}$O gradient in the inner protoplanetary disk. The spatial gradient of the nucleosynthetic isotopic anomaly may be a result of changing isotopic compositions in the infalling materials, or reflect the progressive thermal processing of presolar materials. In contrast, the absence of a $\Delta ^{17}$O gradient reflects that the oxygen isotopic mass-independent fractionation might have altered the spatial distribution of the nucleosynthetic $\Delta ^{17}$O variation in the inner protoplanetary disk before protoplanets formed.}

  \titlerunning{Distribution of isotopes in the inner protoplanetary disk}
  \authorrunning{K. Shuai et al.}
  \maketitle
%

\section{Introduction}

   The isotopic compositions of extraterrestrial samples play a key role in constraining terrestrial planet formation. The mass-independent isotopic variations of meteorites are particularly useful in the classification of meteorites and in tracing their parent bodies \citep[e.g.,][]{Greenwood2020}. A dichotomy of mass-independent isotopic variations has been observed between the noncarbonaceous (NC) meteorites from the inner protoplanetary disk and the carbonaceous (CC) meteorites from the outer disk \citep{Warren2011}. This dichotomy implies a spatial and/or temporal separation between the NC and CC reservoirs \citep{Kruijer2017, Schiller2018, Brasser2020, Johansen2021, Lichtenberg2021, Liu2022}. During the formation of the terrestrial planets, accretion materials from different regions with different isotopic compositions were mixed \citep{Carlson2018}. The compositional diversity of main-belt asteroids also indicates that these asteroids originated from different regions in the Solar System \citep{DeMeo2014}. Therefore, the initial distribution of isotopes and the mixing process occurring during the accretion of planetary bodies in the protoplanetary disk determined the isotopic compositions of the terrestrial planets.
   
\begin{figure}
    \resizebox{\hsize}{!}{\includegraphics{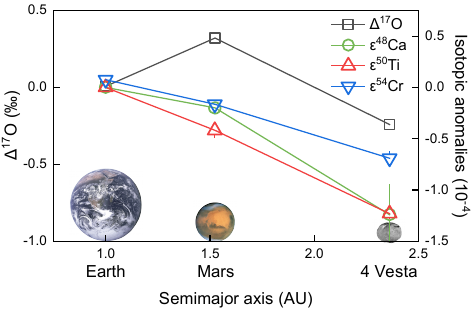}}
    \caption{Mass-independent isotopic variations of Earth, Mars, and Vesta and their current semimajor axes. Error bars smaller than the symbols are not shown. The sizes of the bodies are not to scale. The isotopic data are from the literature \citep[Table~\ref{table:A1};][]{Burkhardt2021}.
    }
        \label{fig:1}
   \end{figure}

   Various mass-independent isotopic signatures have been observed in meteorites. No known chemical process in the solar nebula and planetary bodies could fractionate the nucleosynthetic isotopic anomalies of nonvolatile elements such as $\varepsilon^{54}$Cr, $\varepsilon^{48}$Ca, and $\varepsilon^{50}$Ti \citep{Trinquier2007, Trinquier2009, Qin2010, Dauphas2014, Schiller2015}. These isotopic anomalies reflect the incomplete mixing of stellar nucleosynthetic products \citep{Jacquet2019, Nanne2019} or the unmixing of carriers of these products by physical processes such as thermal processing \citep{Trinquier2009,  VanKooten2016}. In contrast, the isotopic fractionation during CO self-shielding \citep{Yurimoto2004} and symmetry-dependent reactions \citep{Thiemens2021} could have contributed to the mass-independent variation of oxygen isotopes ($\Delta ^{17}$O) in meteorites, in addition to the inherited nucleosynthetic $\Delta ^{17}$O variation. The semimajor axes of the current orbits of Earth, Mars, and Vesta appear to correlate with the isotopic anomalies of terrestrial rocks, Martian meteorites, as well as howardites, eucrites, and diogenites (HED meteorites) from Vesta \citep[Fig.~\ref{fig:1};][]{Yamakawa2010}, whereas no correlation has been observed between their semimajor axes and $\Delta^{17}$O (Fig.~\ref{fig:1}). Notably, planet migration and scattering could have altered the orbits of these three bodies; thus, their current orbits may not represent their feeding zones \citep[e.g.,][]{Bottke2006}. Therefore, it is necessary to constrain the spatial distribution of isotopes in the inner protoplanetary disk.
   
   High-precision isotopic data have been used to constrain or even to calibrate the dynamical models of terrestrial planet formation. The mass-weighted mean initial semimajor axes ($a_{\rm weight}$) of terrestrial planets and other bodies modeled in N-body simulations were used to track the isotopic compositions of accreted materials of planetary bodies \citep{Kaib2015, Mastrobuono-Battisti2015, Fischer2018}. The isotopic difference and the $a_{\rm weight}$ of Earth and Mars simulated in different scenarios indicate that Jupiter and Saturn might have resided on more circular orbits during gas disk dissipation than their current orbits \citep{Woo2021b} and that the terrestrial planets may have accreted locally with limited radial mixing of accretion materials \citep{Mah2021}. In addition, the assumed linear relationship between $a_{\rm weight}$ values and mass-independent isotopic compositions has enabled the estimation of the isotopic compositions of the Moon-forming impactor \citep{Pahlevan2007, Kaib2015, Mastrobuono-Battisti2015} and of the accretion regions of the NC meteorite parent bodies \citep{Shuai2022}. However, the correlation between $a_{\rm weight}$ and isotopic variations relies on a hypothetical spatial isotopic gradient in the protoplanetary disk.
   
   The migration and radial mixing of accretion materials of planets as well as the formation and migration of main-belt asteroids can be modeled by N-body simulations \citep[e.g.,][]{Raymond2017}. Venus, Earth, and Mars can be identified in the N-body simulations of terrestrial planet formation by their orbits and masses \citep[e.g.,][]{Raymond2009,Kaib2015,Fischer2018}. The difference between the isotopic compositions of Earth and Mars can be used to assess the extent of radial mixing in the disk \citep{Fischer2018, Mah2021, Woo2021b}. However, the correlation between $a_{\rm weight}$ and isotopic compositions cannot be verified owing to the lack of isotopic data for Venusian rocks. Nevertheless, differentiated meteorites from main-belt asteroids can represent the bulk mass-independent isotopic compositions of their parent bodies \citep{Greenwood2020}. The isotopic data for differentiated meteorites can be used to calibrate the simulation results and verify the model assumptions on the spatial distribution of isotopes. The comparison between the initial semimajor axes of Vesta analogs in N-body simulations and the trend of isotopic anomalies for Earth, Mars, and Vesta supports the presence of a gradient of isotopic anomalies in the protoplanetary disk \citep{Mah2022b}. In addition to the meteorites from Vesta, isotopic anomalies have been measured for many differentiated meteorites, which represent the bulk isotopic compositions of their parent bodies. On the other hand, the spatial distribution of $\Delta^{17}$O that is different from those of the isotopic anomalies remains poorly understood. Statistical analyses of the isotopic data and the simulated $a_{\rm weight}$ can elucidate the spatial distribution of isotopes in the protoplanetary disk.
   
   The isotopic data for terrestrial planets and meteorites have been interpreted as the result of temporal evolution of isotopic composition in the inner protoplanetary disk \citep{Schiller2018}. The influx of pebbles from the outer Solar System was proposed to result in the change of isotopic compositions in the inner disk. The terrestrial planets and the meteorite parent bodies accreted in different timescales and thus recorded different isotopic compositions of the inner disk. This scenario was further investigated using a pebble accretion model, which reproduced the isotopic compositions of terrestrial planets by fitting the free parameters of the pebble accretion process \citep{Johansen2021}. This temporal evolution model does not exclude the spatial isotopic heterogeneity in the inner protoplanetary disk. During accretion, the pebbles with distinct isotopic compositions could have drifted and mixed. The planetesimals that formed via pebble accretion in the different regions of the inner disk could have different isotopic compositions.
   
   In the present study, we focused on the formation of NC meteorite parent bodies and terrestrial planets in the inner Solar System. We performed N-body simulations to model the isotopic mixing processes during planet accretion and migration. The $a_{\rm weight}$ values of main-belt asteroids and terrestrial planets were modeled in the simulations. The simulated statistical distribution of $a_{\rm weight}$ for the analogs of asteroids and terrestrial planets was compared with the isotopic anomalies and $\Delta^{17}$O of the NC meteorite parent bodies, Earth, and Mars. Our results are consistent with the spatial gradient of isotopic anomalies and the absence of a $\Delta^{17}$O gradient in the inner protoplanetary disk, providing new insights into water transport in the early inner Solar System.
   
\section{N-body simulations} 
   
   We performed simulations of terrestrial planet formation to obtain the $a_{\rm weight}$ of asteroids and terrestrial planets. A GPU-accelerated N-body code GENGA \citep{Grimm2014} was used, which adopts a hybrid symplectic integrator to handle close encounters with good energy conservation. Each simulation starts from a disk with planetesimals, but without any fully formed planetary embryo, similar to the literature studies using GENGA \citep{Hoffmann2017, Woo2021a}. In the simulations, the mutual gravity between planetesimals is included, which requires a large amount of force term calculations. The formation of planetary embryos by accreting planetesimals is simulated, unlike previous simulations that did not consider the mutual gravity between planetesimals \citep[e.g.,][]{Kaib2015, Mastrobuono-Battisti2015}. Thus, the accreted planetesimals were mixed during embryo formation in our simulations. The embryos did not simply have the $a_{\rm weight}$ assigned at the beginning of the simulations. In addition, the inclusion of gravitational interactions between planetesimals enables simulation of the accretion of bodies smaller than embryos; these small bodies can also have mixed $a_{\rm weight}$ values rather than assigned ones. Therefore, our simulations can more realistically model the mixing process during accretion of terrestrial planets and asteroids.
   
   At the beginning of each simulation, equal-mass planetesimals with a radius of 800 km and density of 3 g cm$^{-3}$ were distributed from 0.5 to 3 AU. All planetesimals resided on nearly circular (eccentricity $e$ <0.02), low inclination (inclination $i$ <1$^\circ$) orbits. The other angular orbital elements of these planetesimals were randomly assigned. The gas disk model implemented in GENGA was used, with an e-folding decay timescale of 2 Myr for the gas surface density, which simulates the dissipation of the gas disk in the first 10 Myr during formation of Solar System \citep{Woo2021a}. Circular orbits were adopted for Jupiter and Saturn (CJS) as expected as the initial conditions for the Nice model, and Jupiter and Saturn resided on orbits with a mutual inclination of 0.5$^\circ$ and semimajor axes of 5.45 AU and 8.18 AU, respectively \citep{Tsiganis2005}. In the simulation results of CJS scenario in the previous study, the accretion materials of Earth and Mars were not fully homogenized during the mixing of the disk, consistent with the compositional difference between Earth and Mars \citep{Woo2021b}. Furthermore, our statistical investigation of the distribution of $a_{\rm weight}$ for meteorite parent bodies required a large number of asteroid analogs at the end of the simulations, which could be obtained in the CJS scenario \citep{Woo2022}. The CJS scenario could lead to very massive Mars and main asteroid belt \citep{Raymond2009, Izidoro2014, Woo2022}. However, the early instability of the giant planets may have reduced the mass in the regions where Mars and the main belt formed \citep{Clement2018}. In the present study, we mainly focused on the compositional evolution during the formation of the terrestrial planets and asteroids to investigate the spatial distribution of isotopes in the protoplanetary disk.
   
   \begin{figure}
    \centering
    \includegraphics[width=8 cm]{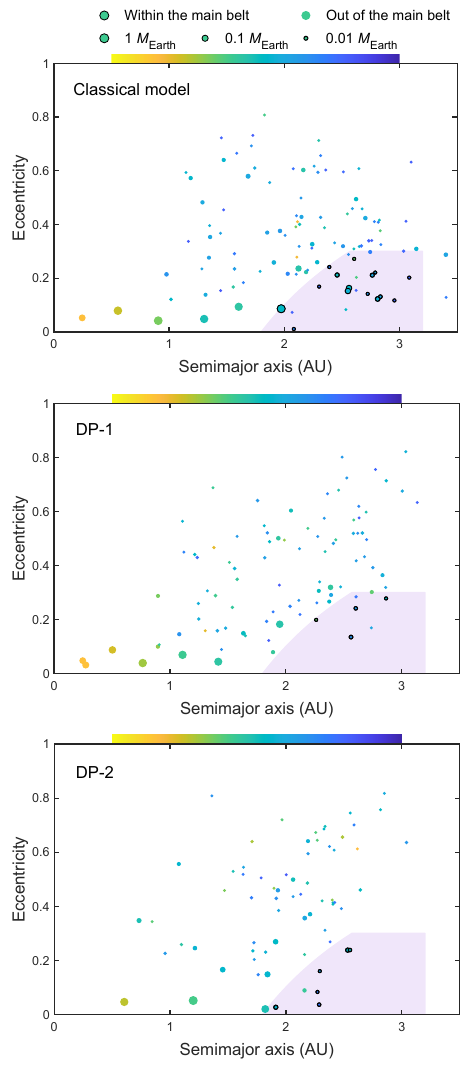}
    \caption{Orbital distributions of bodies at the end of simulations. Typical results in one simulation for each model are shown in each panel. The color coding represents the $a_{\rm weight}$ (average location of feeding zone) of the bodies. The shaded region is the main belt region defined by perihelion distance $q$ >1.8 AU, eccentricity $e$ <0.3, inclination $i$ <25$^\circ$, and semimajor axis $a$ <3.2 AU \citep{Raymond2017}. Each body within the main belt region defined as an asteroid analog is represented by a filled circle with a black outline. The symbol size of each body is proportional to its radius but is not to scale on the $x$ axis.
    }
        \label{fig:2}
   \end{figure}
   
   Various solid surface density profiles were used in previous N-body simulations of terrestrial planet formation. In the classical model \citep{Raymond2009}, the solids distribute in a smooth surface density profile following the minimum-mass Solar Nebula (MMSN) model \citep{Weidenschilling1977, Hayashi1981}. However, the surface density distribution may not have been smooth, and variations in mass flux in different regions of the disk may have resulted in local mass depletion \citep{Jin2008}. This mass depletion has been applied in N-body simulations of the depleted disk model \citep{Izidoro2014, Mah2021, Woo2021a}. Both the classical model and the depleted disk model were numerically simulated in this study. In the classical model, the solid surface density ($\Sigma$) varies with the heliocentric distance $r$ \citep{Weidenschilling1977, Hayashi1981}:
   \begin{linenomath}
       \begin{equation}
       \label{eq:1}
       \Sigma=\Sigma_1(r/1\ \rm{AU})^{-3/2},
       \end{equation}
   \end{linenomath}
   where $\Sigma_1$ = 7 g cm$^{-2}$. The disk consists of 3135 planetesimals with a total mass of $\sim$3.4 $M_{\rm Earth}$ in the classical model. In addition, we performed three high-resolution simulations for the classical model to investigate the effects of the initial number of planetesimals. In these high-resolution simulations, the disk consisted of 12845 planetesimals with a radius of 500 km, and the other parameters remained the same.
   
   Two different sets of parameters of the depleted disk model \citep{Mah2021, Woo2021a} that consider different extents and positions of mass depletion (designated hereafter as DP-1 and DP-2, respectively) were used:
   \begin{linenomath}
       \begin{equation}
       \label{eq:2}
       \Sigma = \left \{
       \begin{aligned}
           &\chi\Sigma_1(r/1\ \rm{AU})^{-3/2}; & r \leq r_{\rm dep}, \\
           &(1-\beta)\chi\Sigma_1(r/1\ \rm{AU})^{-3/2}; & r > r_{\rm dep}.
       \end{aligned} 
       \right.
       \end{equation}
   \end{linenomath}
   In DP-1, $\chi$ = 1, $\beta$ = 0.5, and $r_{\rm dep}$ = 1.5 AU; the disk consisted of 2360 planetesimals with a total mass of $\sim$2.5 $M_{\rm Earth}$. In DP-2, $\chi$ = 2, $\beta$ = 0.75, and $r_{\rm dep}$ = 1 AU; the disk consisted of 2912 planetesimals with a total mass of $\sim$3.1 $M_{\rm Earth}$. The compositional mixing process in the CJS scenario with a depleted disk has not been investigated previously. Therefore, we performed CJS simulations with a depleted disk to investigate whether the compositional difference between Earth and Mars could be reproduced with the mass depletion and, more importantly, whether the distribution of isotopic compositions among the meteorite parent bodies could be reproduced.

   We performed ten simulations for each model (the classical model, DP-1, and DP-2) for 150 Myr with a time step of 6 days. The giant impacts could mix different isotopic compositions \citep{Kaib2015, Mastrobuono-Battisti2015} and thus the late giant impact stage lasting until $\sim$150 Myr after Solar System formation \citep{Touboul2007} was included in the simulations. All collisions were assumed to be inelastic and conserve linear momentum. The mass-weighted mean initial semimajor axis ($a_{\rm weight}$) of each planetary body at the end of the simulations was calculated as the average location of its feeding zone \citep{Kaib2015, Mastrobuono-Battisti2015, Fischer2018, Mah2021}:
   \begin{linenomath}
       \begin{equation}
       \label{eq:3}
       a_{\rm weight}=\frac{\Sigma^N_im_ia_i}{\Sigma^N_im_i},
       \end{equation}
   \end{linenomath}
   where $m_i$ and $a_i$ are the initial mass and the initial semimajor axis of the $i$th planetesimal accreted into the planetary body, and $N$ is the number of accreted planetesimals. The mass-independent isotopic compositions of the accretion materials were homogenized in each differentiated body. Therefore, the width of a body’s feeding zone could not be recovered in its meteorites, and the width of the feeding zone was not used in this study.

\section{Results}
\begin{figure}
    \centering
    \includegraphics[width=8 cm]{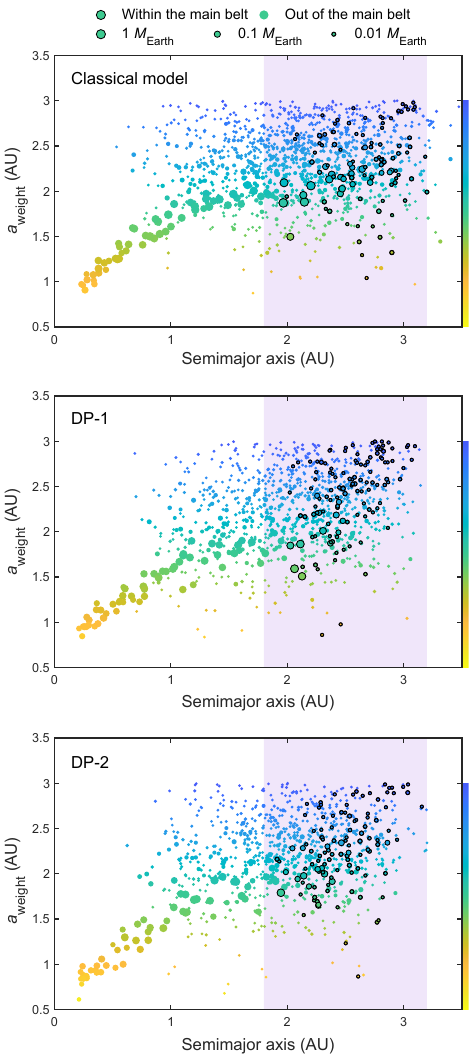}
    \caption{Comparison between the feeding zones and the final semimajor axes of bodies at the end of all simulations. All simulations for each model are included. The color coding represents the $a_{\rm weight}$ (average location of feeding zone) of the bodies. The shaded region is the main belt region defined by perihelion distance $q$ >1.8 AU, eccentricity $e$ <0.3, inclination $i$ <25$^\circ$, and semimajor axis $a$ <3.2 AU \citep{Raymond2017}. Each body within the main belt region defined as an asteroid analog is represented by a filled circle with a black outline. The symbol size of each body is proportional to its radius but is not to scale on the $x$ and $y$ axes.
    }
        \label{fig:3}
   \end{figure}

\begin{figure}
    \centering
    \includegraphics[width=7.8 cm]{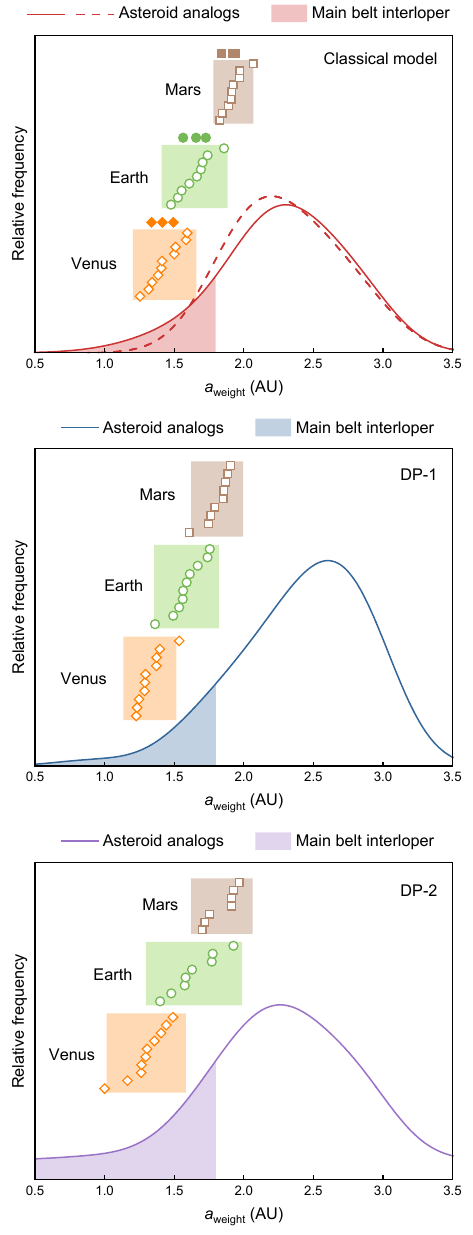}
    \caption{Average feeding zone locations of the analogs of the terrestrial planets and main-belt asteroids in the simulations. The open symbols are the $a_{\rm weight}$ values of the terrestrial planet analogs. The shaded bands represent 2$\sigma$ of $a_{\rm weight}$ for each terrestrial planet analog. The solid curves are the kernel density estimates of the $a_{\rm weight}$ values of main-belt asteroid analogs. The shaded areas under the curves represent the portion of main-belt interlopers ($a_{\rm weight}$ <1.8 AU) in the main-belt asteroid analogs. The filled symbols and the dashed curve are the $a_{\rm weight}$ values of the terrestrial planet analogs and the kernel density estimate of $a_{\rm weight}$ values of main-belt asteroid analogs in the high-resolution simulations, respectively.
    }
        \label{fig:4}
   \end{figure}

   The simulation results for the classical model and for DP-1 and DP-2 are similar (Fig.~\ref{fig:2}). The final orbits of planetary bodies are consistent with the results of previous CJS simulations \citep{Hoffmann2017, Woo2022}. The $a_{\rm weight}$ values of the planetary bodies calculated using Eq. (\ref{eq:3}) range from 0.5 to 3 AU (Fig.~\ref{fig:3}). Planets and embryos are defined as bodies larger than $3.3\times10^{23}$ kg (0.055 $M_{\rm Earth}$), and smaller bodies are considered planetesimals \citep{Hoffmann2017}. In the simulation results for all models, the planets and embryos have relatively low eccentricities, with an average of 0.07 $\pm$ 0.05, whereas the orbits of the remaining planetesimals are excited by the gas giants, planets, and embryos, with an average eccentricity of 0.38 $\pm$ 0.20 (Fig.~\ref{fig:2}). The analogs of Venus, Earth, and Mars are identified as the largest planets or embryos in the regions of 0.5 AU <$a$ <0.85 AU, 0.85 AU <$a$ <1.25 AU, and 1.25 AU <$a$ <1.75 AU \citep{Woo2021b}, respectively, where $a$ is the final semimajor axis of the planetary body. The feeding zones of the terrestrial planets are listed in Table~\ref{table:1}. Venus, Earth, and Mars show statistically different feeding zones (Fig.~\ref{fig:4}). The feeding zones of the terrestrial planets in the classical model are in accord with those reported in the literature within uncertainties \citep{Fischer2018}.

   \begin{table}
    \caption{Average feeding zone locations ($a_{\rm weight}$) of the terrestrial planet and main-belt asteroid analogs in the simulations.}
        \label{table:1}
    \centering
    \begin{tabular}{l c c c}
        \hline\hline
        & \multicolumn{3}{c}{Average feeding zone location (AU)} \\    
        & Classical model & DP-1 & DP-2 \\
        \hline                        
        Venus analog & 1.4 $\pm$ 0.2 & 1.3 $\pm$ 0.2 & 1.3 $\pm$ 0.3 \\      
        Earth analog & 1.6 $\pm$ 0.2 & 1.6 $\pm$ 0.2    & 1.6 $\pm$ 0.3 \\
        Mars analog & 1.9 $\pm$ 0.1 & 1.8 $\pm$ 0.2     & 1.8 $\pm$ 0.2 \\
        Asteroid analogs\tablefootmark{a} & 2.3 $\pm$ 0.8 & 2.4 $\pm$ 0.9    & 2.3 $\pm$ 0.8 \\
        \begin{tabular}{@{}l@{}}Main-belt \\ interlopers\tablefootmark{a,b}
    \end{tabular} 
    & 1.5 $\pm$ 0.4 & 1.5 $\pm$ 0.6    & 1.6 $\pm$ 0.5 \\ 
    \hline                                   
    \end{tabular}
    \tablefoot{
   \tablefoottext{a}{The values are the averages of $a_{\rm weight}$ and their respective 2$\sigma$ for all asteroid analogs or main-belt interlopers in the simulations of each model.} \\
   \tablefoottext{b}{Main-belt interlopers are asteroid analogs with $a_{\rm weight}$ <1.8 AU.}
   }
   \end{table}
   
   The analogs of main-belt asteroids in the simulations are bodies within the main asteroid belt (perihelion distance $q$ >1.8 AU, $e$ <0.3, $i$ <25$^\circ$, and $a$ <3.2 AU) \citep{Raymond2017}. The number of asteroid analogs that formed in each low-resolution simulation varies from 4 to 34, with averages of 11, 12, and 11 for the classical model, DP-1, and DP-2, respectively. An average of 18 asteroid analogs formed in each high-resolution simulation. The cumulative distributions of $a_{\rm weight}$ of asteroid analogs in different simulations of each model are similar, particularly for simulations in which at least ten asteroid analogs formed (Fig.~\ref{fig:5}). The $a_{\rm weight}$ values of main-belt asteroid analogs in each model vary from $\sim$1 AU to 3 AU (Fig.~\ref{fig:3}). The lower limit of this range in the simulations is much closer to the Sun than the lower boundary ($\sim$2 AU) of the main belt region, suggesting that a substantial number of asteroids in the main belt originated from outside the main belt. We define planetesimals that are in the main belt, but have $a_{\rm weight}$ <1.8 AU at the end of simulations as main-belt interlopers, which were also observed in previous studies \citep{Bottke2006, Raymond2017}. In our simulations, 11\%, 8\%, and 4\% of the asteroid analogs are main-belt interlopers in the classical model, DP-1, and DP-2, respectively (Fig.~\ref{fig:4}). A few asteroid analogs have $a_{\rm weight}$ values comparable to or even smaller than that of Earth ($\sim$1.6 AU).
\section{Discussion}
   \subsection{Feeding zones of terrestrial planets and asteroids}
   The correlations of $a_{\rm weight}$ and final semimajor axis ($a$) are markedly different between the planets (and embryos) and the remaining planetesimals in the simulation results (Fig.~\ref{fig:3}). In all of the models, the $a_{\rm weight}$ values of the planets and embryos correlate with their final semimajor axes (Fig.~\ref{fig:3}), indicating that the initial spatial distribution of the isotopic compositions is preserved in these bodies. The radial mixing during the collisional accretion of planets and embryos is not sufficiently intense to fully homogenize their accretion materials (planetesimals) from different regions, consistent with findings from previous CJS simulations \citep{Woo2021b}. Intriguingly, the $a_{\rm weight}$ values of the planets and embryos are generally larger than their final semimajor axes, particularly for those in the terrestrial planet region ($a$ <2 AU) (Fig.~\ref{fig:3}). These larger $a_{\rm weight}$ values indicate that, on average, the planets and embryos accreted planetesimals from regions further away from the Sun. The inward migration and accretion of planetesimals resulted from hydrodynamic drag by the gas disk before the nebular gas dissipation in the first 10 Myr of Solar System formation and from the orbital resonances of the Jupiter–Saturn system \citep{Hoffmann2017}. The inner edge of the planetesimal disk shaped by the piling up of materials \citep{Youdin2004}, is at 0.5 AU in our simulations, consistent with those in previous studies \citep[e.g.,][]{Raymond2009,Kaib2015,Mah2021,Woo2021a}. The uncertainty of the location of this inner edge could affect the accretion of Mercury, which is not included in our discussion. By contrast, the other terrestrial planets mainly accumulated materials in regions further out than their current orbits. Therefore, the cutoff at the inner edge of the planetesimal disk only has a minor effect on the $a_{\rm weight}$ values of Venus, Earth, and Mars.
   
   The remaining planetesimals with similar final semimajor axes show large variations in their $a_{\rm weight}$ values, and no correlation between $a_{\rm weight}$ and $a$ for planetesimals is observed (Fig.~\ref{fig:3}). Even in the region of >2 AU, remaining planetesimals with $a_{\rm weight}$ <1 AU is present, reflecting their large-scale outward migration. Therefore, the migration of the planetesimals disturbed their initial spatial distribution. The feeding zones of both terrestrial planets and asteroids modeled in the high-resolution simulations are in line with those modeled in the low-resolution simulations (Fig.~\ref{fig:4}). Therefore, the initial number of planetesimals did not significantly affect the isotopic mixing process in the simulations.

   \begin{figure}
    \centering
    \includegraphics[width=7.5 cm]{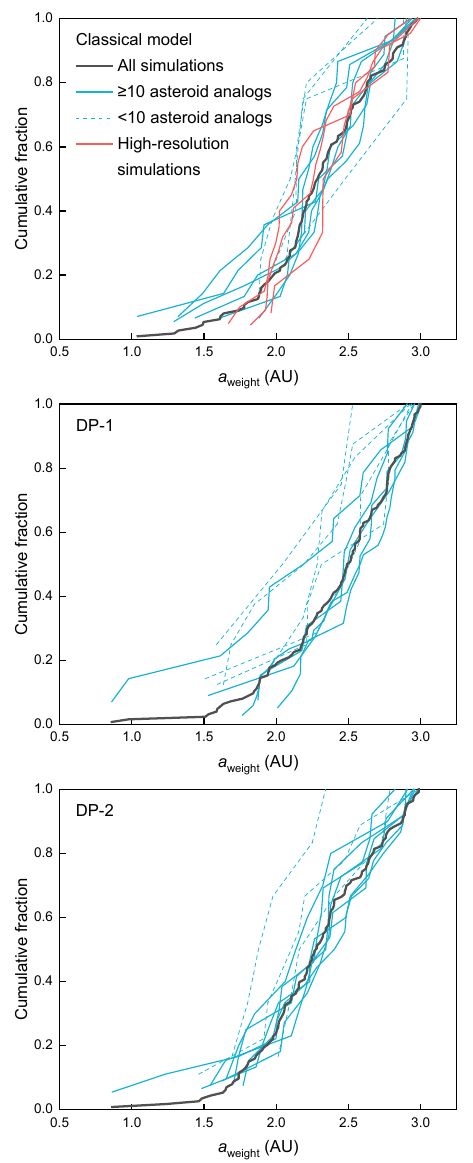}
    \caption{Cumulative distribution of $a_{\rm weight}$ for main-belt asteroid analogs in the simulations. The thick black lines represent the $a_{\rm weight}$ distributions for asteroid analogs in all simulations of each model. The solid and dashed blue lines correspond to the $a_{\rm weight}$ distributions for asteroid analogs in a single simulation with $\geq$10 asteroid analogs and <10 asteroid analogs, respectively. The solid red lines represent the $a_{\rm weight}$ distributions for asteroid analogs in each high-resolution simulation of the classical model.
    }
        \label{fig:5}
   \end{figure}
   \subsubsection{Feeding zones of the terrestrial planets}
   The analogs of Earth, Mars, and Venus have statistically different feeding zones in the classical model, DP-1, and DP-2 (Fig.~\ref{fig:4}). The accretion materials in the region where the terrestrial planets formed are not fully homogenized, consistent with the isotopic difference between Earth and Mars (Fig.~\ref{fig:1}). The feeding zones of the terrestrial planets in DP-1 are statistically closer to the Sun than those in the classical model (Fig.~\ref{fig:4}). This difference indicates that the mass depletion in the >1.5 AU region affected the accretion of Earth, Mars, and Venus, which was also observed in the previous studies of the depleted disk model \citep{Izidoro2014, Mah2021}. However, the difference between these models for each terrestrial planet is smaller than the uncertainties.
   
   The feeding zones of each terrestrial planet in the ten simulations for each model show $\sim$0.5 AU variation (Fig.~\ref{fig:4}). This variation reflects the stochasticity of terrestrial planet formation, indicating that the feeding zones of the terrestrial planets are difficult to be determined precisely. Furthermore, the feeding zones of Earth and Mars differ by only $\sim$0.2 AU, which is comparable in value to the uncertainties of the feeding zones of Earth and Mars (Table~\ref{table:1}). This $\sim$0.2 AU difference is substantially smaller than the range of feeding zones of the other bodies (Fig.~\ref{fig:3}). Therefore, extrapolation of the feeding zones of Earth and Mars to other bodies can lead to large uncertainty, typically 0.2–0.7 AU \citep{Shuai2022}.

   \subsubsection{Feeding zones of asteroids}
   The main-belt asteroids formed in the inner Solar System have been considered as the parent bodies of NC meteorites \citep{Greenwood2020}. It is difficult to link an asteroid analog in our simulation to a specific meteorite parent body. However, assuming the asteroid analogs in our simulations are intact bodies rather than rubble piles, each asteroid analog can represent an NC meteorite parent body that has been defined based on the isotopic compositions of the NC meteorites \citep[Table~\ref{table:A1};][]{Greenwood2020}. The $a_{\rm weight}$ distribution of the asteroid analogs can be statistically compared to the distribution of isotopic compositions of the NC meteorite parent bodies.
   
   In different simulations of each model, the cumulative distributions of the $a_{\rm weight}$ values of the asteroid analogs are similar, particularly for simulations with no less than ten asteroid analogs (Fig.~\ref{fig:5}). The two-sample Kolmogorov–Smirnov (K–S) test is used to quantitatively compare the cumulative distributions of $a_{\rm weight}$ of asteroid analogs. If the K–S test yields a probability $p$ <0.05, the two compared datasets have distinct distributions; otherwise, the two datasets have similar distributions \citep[e.g.,][]{Mastrobuono-Battisti2015}. We use the K–S test to compare the $a_{\rm weight}$ distribution of the asteroid analogs in each simulation with that of all the asteroid analogs in the other nine simulations of the same model. The K–S test yields $p$ >0.05 for all simulations, indicating that the $a_{\rm weight}$ values of asteroid analogs in different simulations of the same model have similar distributions. In addition, the most common $a_{\rm weight}$ of all asteroid analogs in DP-1 is $\sim$2.6 AU, which differed from that in the classical model ($\sim$2.3 AU) (Fig.~\ref{fig:4}). The K–S test of the $a_{\rm weight}$ distributions between the classical model and DP-1 yields $p$ = 0.01, suggesting that these two distributions in different models are statistically distinct. In contrast, the $a_{\rm weight}$ distribution of all asteroid analogs in DP-2 is similar to that in the classical model, with $p$ = 0.61. There are more asteroid analogs with low $a_{\rm weight}$ values in DP-2 than DP-1, which might correlate with larger solid surface density resulting from larger $\chi$ in Equation (2) in DP-2 than that in DP-1. The $a_{\rm weight}$ distributions of the asteroid analogs in the high-resolution simulations of the classical model are in line with those in the low-resolution simulations (Figs. \ref{fig:4} and \ref{fig:5}).
   
   In the simulation results for all of the models, the asteroid analogs exhibit large ranges for $a_{\rm weight}$ from $\sim$1 AU to 3 AU (Fig.~\ref{fig:3}), indicating that not all asteroids formed in situ in the main belt zone. Some planetesimals originating outside the main belt zone were scattered into this zone. These main-belt interlopers formed in the terrestrial planet region and accreted materials similar to the building blocks of the terrestrial planets \citep{Shuai2022}. Enstatite chondrites and aubrites are isotopically similar to Earth (Table~\ref{table:A1}), and their parent bodies are likely to reside in the main asteroid belt \citep{Greenwood2020}, consistent with the presence of these main-belt interlopers. In summary, the $a_{\rm weight}$ values of asteroid analogs have wide ranges, and their distributions in different simulations for the same model are similar. Therefore, the $a_{\rm weight}$ distribution of asteroid analogs is useful in calibrating the simulation results.
   
   \subsection{Isotopic compositions of NC meteorite parent bodies}
   The distribution of the mass-independent isotopic data for the NC meteorite parent bodies is compared with the simulated $a_{\rm weight}$ distribution of the asteroid analogs. Only meteorites with isotopic compositions within the range of NC meteorites \citep{Warren2011} or those related to other NC meteorites \citep{Greenwood2020} are included. The isotopic compositions of the NC meteorite parent bodies are calculated as the weighted average of the isotopic compositions of the meteorites (Table~\ref{table:A1}). We use the isotopic data for the meteorites that can be linked to their parent bodies. These meteorites include magmatic achondrites, primitive achondrites, and stony-iron meteorites. The mass-independent isotopic compositions of magmatic achondrites and stony-iron meteorites can represent those of their parent bodies because the isotopic compositions have been homogenized in each parent body during global melting \citep{Greenwood2017}. Therefore, the isotopic compositions of magmatic achondrites and stony-iron meteorites are used in the present study. The primitive achondrites experienced less extensive melting than the magmatic achondrites and show a range of isotopic variation in each primitive achondrite group (Table~\ref{table:A1}); however, their parent bodies are relatively well defined \citep[4-5 parent bodies;][]{Greenwood2020}. Therefore, the primitive achondrites are also used in this study. In contrast, the number of parent bodies of iron meteorites is unclear \citep[26–60 parent bodies;][]{Greenwood2020}. The NC chondrites, including ordinary chondrites, enstatite chondrites, and Rumuruti chondrites, are undifferentiated and show relatively large $\Delta^{17}$O variation in each group compared with the magmatic achondrites \citep[Table~\ref{table:A1};][]{Greenwood2017}. The number of parent bodies for NC chondrites is uncertain \citep[5–10 parent bodies;][]{Greenwood2020}; therefore, the isotopic data for iron meteorites and NC chondrites are not used. Nevertheless, the isotopic data of NC chondrites do not significantly affect the distribution of isotopic data for the NC meteorite parent bodies and the conclusion is not changed.
   
   \begin{figure}
    \centering
    \includegraphics[width=7.5 cm]{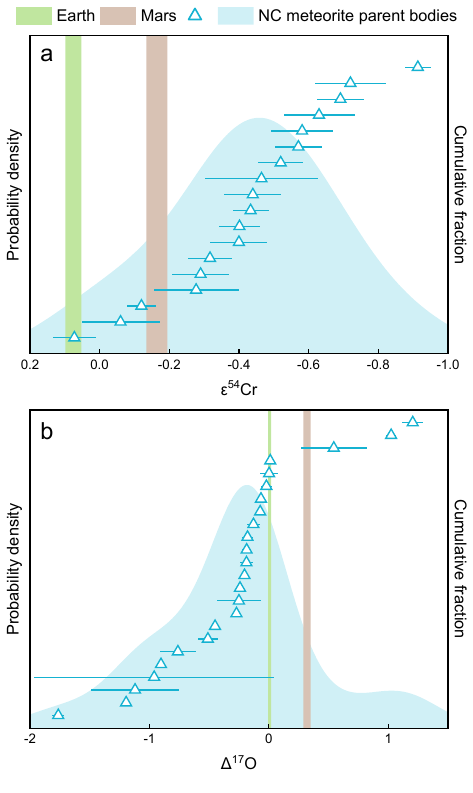}
    \caption{Mass-independent isotopic data for Earth, Mars, and NC meteorite parent bodies. The $\varepsilon^{54}$Cr data are reversely plotted for ease of comparison with the $\Delta^{17}$O data and simulation results. The blue triangles are the $\varepsilon^{54}$Cr and $\Delta^{17}$O data for the NC meteorite parent bodies, shown as the cumulative distribution of these data with the $y$ axis representing the cumulative fraction. The blue-shaded areas are the kernel density estimates of these data with the $y$ axis representing the probability density. The green and brown bands represent the $\varepsilon^{54}$Cr and $\Delta^{17}$O values for Earth and Mars with uncertainties, respectively. The isotopic data are from the literature (Table~\ref{table:A1}).
    }
        \label{fig:6}
   \end{figure}
   
   The isotopic anomalies of different elements such as $\varepsilon^{48}$Ca, $\varepsilon^{50}$Ti, and $\varepsilon^{54}$Cr in NC meteorites are correlated \citep{Trinquier2009, Schiller2015}. We use $\varepsilon^{54}$Cr to represent the distribution of isotopic anomalies in the NC meteorite parent bodies because $\varepsilon^{54}$Cr has also been measured in the ungrouped achondrites and in anomalous basaltic meteorites \citep[e.g.,][Table~\ref{table:A1}]{Wimpenny2019}. These meteorites represent separate parent bodies and provide additional data to the NC parent bodies \citep{Greenwood2020}. The distributions of $\varepsilon^{54}$Cr and $\Delta^{17}$O in NC meteorite parent bodies (Table~\ref{table:A1}) are compared with the $\varepsilon^{54}$Cr and $\Delta^{17}$O values for Earth and Mars (Fig.~\ref{fig:6}). The $\varepsilon^{54}$Cr data for Earth, Mars, and the NC meteorite parent bodies (Fig.~\ref{fig:6}a) show a similar distribution to the $a_{\rm weight}$ distributions simulated in the classical model, DP-1, and DP-2 (Fig.~\ref{fig:4}). Earth has the same $\varepsilon^{54}$Cr value as main-group aubrites, which exhibit the highest $\varepsilon^{54}$Cr among the NC meteorite parent bodies. Most of the NC meteorite parent bodies have lower $\varepsilon^{54}$Cr values than Earth and Mars, and the most common value is $\sim$–0.5. In contrast to the $\varepsilon^{54}$Cr distribution, the $\Delta^{17}$O distribution of Earth, Mars, and the NC meteorite parent bodies (Fig.~\ref{fig:6}b) differs from the simulated $a_{\rm weight}$ distribution (Fig.~\ref{fig:4}). Earth’s $\Delta^{17}$O does not represent an endmember for the NC meteorite parent bodies but is close to the most common $\Delta^{17}$O value of NC meteorite parent bodies (Fig.~\ref{fig:6}b). Mars has a higher $\Delta^{17}$O value than Earth and most of the NC meteorite parent bodies. These different distributions of $\varepsilon^{54}$Cr and $\Delta^{17}$O are consistent with the different relationships between heliocentric distance and $\varepsilon^{54}$Cr or $\Delta^{17}$O of Earth, Mars, and Vesta (Fig.~\ref{fig:1}) and with the lack of correlation between isotopic anomalies and $\Delta^{17}$O among the NC meteorites \citep{Shuai2022}.
   
   \subsection{Spatial isotopic gradient in the protoplanetary disk}
   The $a_{\rm weight}$ distributions of asteroid analogs in the classical model, DP-1, and DP-2 (Fig.~\ref{fig:4}) are similar to the $\varepsilon^{54}$Cr distribution of NC meteorite parent bodies but appear to differ from the $\Delta^{17}$O distribution of NC meteorite parent bodies (Fig.~\ref{fig:6}). To quantitatively compare the simulated $a_{\rm weight}$ and the isotopic data, we assume a spatial isotopic gradient in the disk to calibrate the simulation results. In this calibration, isotopic compositions are calculated using the $a_{\rm weight}$ values modeled in our simulations. The simplest isotopic gradient—a linear relationship between $a_i$ (initial semimajor axis of the ith planetesimal) and $\varepsilon^{54}$Cr or $\Delta^{17}$O—is used, similar to previous studies \citep{Pahlevan2007, Kaib2015, Mastrobuono-Battisti2015}:
   \begin{linenomath}
       \begin{equation}
           \label{eq:4}
           \varepsilon_i=c_1a_i+c_2
       \end{equation}
   \end{linenomath}
   where $\varepsilon_i$ is the initial $\varepsilon^{54}$Cr or $\Delta^{17}$O of the $i$th planetesimal; $c_1$ and $c_2$ are the parameters of the linear function calculated in the calibration. Both Earth and enstatite chondrites have $\varepsilon^{54}$Cr and $\Delta^{17}$O close to 0, whereas Mars and ordinary chondrites have positive $\Delta^{17}$O and negative $\varepsilon^{54}$Cr. The feeding zone of Mars is generally farther from the Sun than that of Earth both in our results (Fig.~\ref{fig:4}) and in previous simulations \citep[e.g.,][]{Kaib2015, Fischer2018}. A variety of evidence including oxidation state and elemental abundance indicates that the parent bodies of ordinary chondrites formed further away from the Sun than those of enstatite chondrites \citep{Rubie2015}. Therefore, the parameter $c_1$ is positive for $\Delta^{17}$O and negative for $\varepsilon^{54}$Cr, assuming isotopic gradients in the inner protoplanetary disk.
   
   \begin{figure}
    \centering
    \includegraphics[width=7.5 cm]{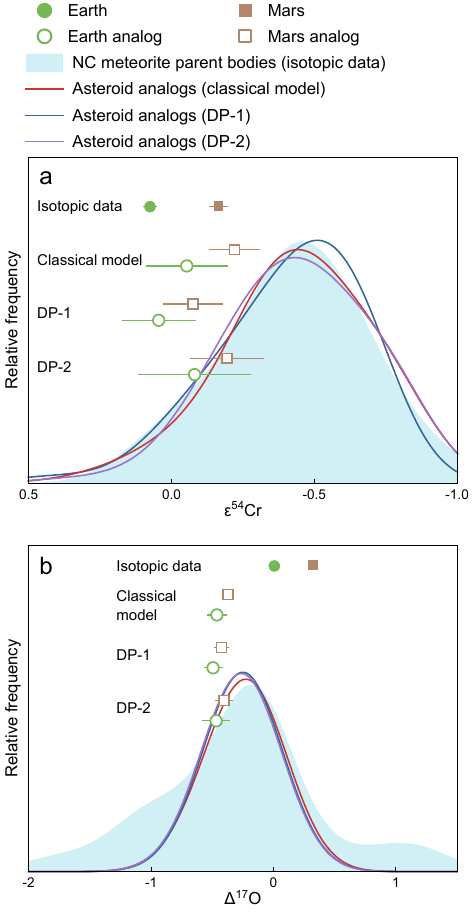}
    \caption{Comparison between the calculated isotopic compositions of simulated analogs and the published isotopic data. The isotopic compositions of the analogs of Earth, Mars, and main-belt asteroids are calculated using Eq. (\ref{eq:4}). The parameters $c_1$ and $c_2$ are chosen for the best match between the calculated isotopic compositions of the asteroid analogs (blue, red, and purple curves) and the published isotopic data for NC meteorite parent bodies (blue shaded area). The best match has the largest $p$ in the Kolmogorov–Smirnov test. All uncertainties are 2$\sigma$ except for the uncertainties of the published $\varepsilon^{54}$Cr data, which are shown as 2se.
    }
        \label{fig:7}
   \end{figure}
   
   The isotopic compositions of each initial planetesimal in our simulations can be calculated using $c_1$, $c_2$, and its initial semimajor axis assuming an isotopic gradient in the inner solar disk. The analogs of asteroids and terrestrial planets formed via accretion of these planetesimals, and the $a_{\rm weight}$ values were calculated using the initial semimajor axes of the initial planetesimals. Therefore, the assumed isotopic gradient applies to the analogs of asteroids and terrestrial planets. We first used the $a_{\rm weight}$ values of asteroid analogs and the isotopic data of meteorites to calculate $c_1$ and $c_2$, that is, to calibrate the assumed isotopic gradient. Then, we used the $a_{\rm weight}$ values and the isotopic data of Earth and Mars to validate the calibrated isotopic gradient.
   
   An optimization routine that calculates parameters $c_1$ and $c_2$ in Equation (4) were adopted to calibrate the isotopic gradient. In this routine, 10$^5$ sets of randomly generated values for $c_1$ and $c_2$ were used to calculate the isotopic compositions of the asteroid analogs using the simulated $a_{\rm weight}$ values and Equation (4). The calculated statistical distribution of the isotopic compositions of the asteroid analogs was compared with the distribution of the isotopic data for NC meteorite parent bodies using the two-sample K–S test. The slope ($c_1$) and the intercept ($c_2$) in Equation (4) were determined by matching the width and the position of the calculated distribution to those of the isotopic data distribution (Fig. 7), which have the largest $p$ value in the two-sample K–S test. The best modeled $c_1$ and $c_2$ for each model (i.e., the classical model, DP-1, or DP-2) and for each isotopic composition ($\varepsilon^{54}$Cr or $\Delta^{17}$O) were calculated. The isotopic compositions of terrestrial planet analogs can be calculated using $c_1$, $c_2$, and the simulated $a_{\rm weight}$ values. The Equation (4) is validated by comparing the calculated results and the published isotopic data of Earth and Mars. 
   
   The calculated $\varepsilon^{54}$Cr values of Earth and Mars analogs are in accord with the published $\varepsilon^{54}$Cr data within uncertainties for all three models (Fig.~\ref{fig:7}a). The best matches between the $\varepsilon^{54}$Cr calculated for asteroid analogs and the $\varepsilon^{54}$Cr published for NC meteorite parent bodies yield $p$ values >0.99 for all of the models. The consistency between our simulations and the observed $\varepsilon^{54}$Cr distributions validates the assumption of a spatial $\varepsilon^{54}$Cr gradient in the inner protoplanetary disk. Our results indicate that the spatial $\varepsilon^{54}$Cr gradient in the inner protoplanetary disk can be approximated using a linear function:
   \begin{linenomath}
       \begin{equation}
           \label{eq:5}
           \varepsilon^{54}{\rm Cr}_i=-0.57a_i+0.90,
       \end{equation}
   \end{linenomath}
   where the parameters are the mean modeled values of the similar $c_1$ (–0.59, –0.55, and –0.57) and $c_2$ (0.91, 0.92, and 0.86) in the classical model, DP-1, and DP-2. The uncertainties of the parameters $c_1$ and $c_2$ are 0.07 and 0.16, respectively, which are the standard deviations of the $c_1$ and $c_2$ values with $p$ values larger than 0.05 in the optimization routine. Isotopic anomalies such as $\varepsilon^{48}$Ca, $\varepsilon^{50}$Ti, and $\varepsilon^{94}$Mo correlate with $\varepsilon^{54}$Cr in NC meteorites \citep{Trinquier2009, Schiller2015, Burkhardt2021}; therefore, spatial gradients were present for these isotopic anomalies as well.
   
   Our results indicate that the collisional accretion process produced the distribution of isotopic anomalies among Earth, Mars, and the NC meteorite parent bodies. The majority of planetesimals accreted by Earth and Mars originated from the <2 AU region. Consequently, Earth and Mars have higher $\varepsilon^{54}$Cr values than most of the NC meteorite parent bodies. The NC meteorite parent bodies consist of planetesimals from a wide range of heliocentric distances, resulting in a wide distribution of $\varepsilon^{54}$Cr, with the most common value of $\sim$–0.5 (Fig.~\ref{fig:7}). Using the quantified $\varepsilon^{54}$Cr gradient of Eq. (\ref{eq:5}), the $\varepsilon^{54}$Cr of Venus can be estimated according to the average $a_{\rm weight}$ of Venus analogs obtained in our simulations (Table~\ref{table:1}). The classical model, DP-1, and DP-2 yield $\varepsilon^{54}$Cr of 0.08 $\pm$ 0.13, 0.14 $\pm$ 0.11, and 0.16 $\pm$ 0.16 for Venus, respectively, which are not distinguishable from the $\varepsilon^{54}$Cr of Earth and are consistent with the estimated values in the literature \citep[0.10 and 0.07;][]{Mah2021}.

   During accretion of terrestrial planets and asteroids, different elemental concentrations in planetesimals might have affected the mixing of isotopic anomalies \citep[e.g.,][]{Fitoussi2016, Dauphas2017, Shuai2022}. However, the bulk concentrations of nonvolatile elements including Cr in planetesimals are generally chondritic, that is, similar to the average composition of the Solar System \citep[e.g.,][]{Toplis2013, Fitoussi2016}. In a two-endmember mixing model of terrestrial planets and NC meteorites that combines elemental concentrations and isotopic anomalies, the $\varepsilon^{54}$Cr values of the two endmembers show large variation from –1.03 to 0.24, but the Cr concentrations of the two endmembers only vary moderately from 0.32 to 0.45 wt.\% \citep{Shuai2022}. Therefore, the difference in Cr concentrations of planetesimals may only have a minor effect on the mixing of Cr isotopes during accretion. In addition, the spatial distribution of Cr concentration in the protoplanetary disk is unknown, as well as the bulk Cr concentrations in most differentiated parent bodies. Therefore, variation of Cr concentrations in different planetesimals is not considered in our model.
   
   In contrast to $\varepsilon^{54}$Cr, the calculated $\Delta^{17}$O values of Earth, Mars, and the main-belt asteroid analogs deviate from the published $\Delta^{17}$O data (Fig.~\ref{fig:7}b). Even though the modeled parameters $c_1$ and $c_2$ are chosen for the largest $p$ in the K–S test, the $\Delta^{17}$O distribution calculated for the asteroid analogs does not match the observed $\Delta^{17}$O data for the NC meteorite parent bodies well. The largest $p$ values for the classical model, DP-1, and DP-2 are 0.13, 0.09, and 0.13, respectively. Negative $\Delta^{17}$O values are calculated for Earth and Mars analogs according to the modeled parameters $c_1$ and $c_2$, inconsistent with the published $\Delta^{17}$O data. Therefore, the assumption that $\Delta^{17}$O varied systematically with heliocentric distance in the inner protoplanetary disk is not valid.
   
   A step function of spatial distribution of isotopes in the inner Solar System has also been adopted in previous studies \citep{Kaib2015, Fischer2018}. This step function, which has a sharp contrast of isotopic compositions in the disk according to the difference in isotopic compositions between different chondritic groups, assumes that the NC chondrites are the building blocks of the planets and asteroids. However, the ranges of nucleosynthetic isotopic anomalies and $\Delta^{17}$O in NC chondrites are smaller than those in all NC meteorites \citep[Table~\ref{table:A1};][]{Shuai2022}. The addition of CC chondrites with positive $\varepsilon^{54}$Cr values cannot account for the $\varepsilon^{54}$Cr values of some NC meteorites (e.g., ureilites) that are lower than those of the NC chondrites. In addition, chondritic materials may represent the crust of differentiated bodies \citep{Elkins-Tanton2011}, which may be derived from the collisional recycling of planetesimals \citep{Lichtenberg2018}. Furthermore, the elemental ratios of NC chondrites do not match those of the terrestrial planets \citep{Drake2002, Dauphas2015}. Therefore, the NC chondrites cannot be the building blocks of planets and asteroids in our simulations. On the other hand, the isotopic compositions of the NC differentiated meteorites vary continuously (Fig.~\ref{fig:6}). The step-function of the spatial distribution of isotopes is not considered in the present study.
   
   In our simulations, Jupiter and Saturn initially reside on circular orbits. The Mars analogs in our simulations are too large in size (mass > 0.2 $M_{\rm Earth}$) (Fig.~\ref{fig:B1}) and their accretion does not complete until 10 Myr after formation of the Solar System, similar to the CJS simulations in the previous studies \citep{Izidoro2014, Hoffmann2017, Woo2022}. The consistency in isotopic compositions and the inconsistencies in mass and accretion time in the CJS simulations indicate that an external agent is required to disturb the system \citep{Woo2022}. The instability of giant planets could have stunted the growth of Mars, leading to the small size and short accretion time of Mars \citep{Clement2018}. The instability time of the giant planets cannot be later than 80 Myr after formation of the Solar System \citep{Mojzsis2019}. The effects of giant planet instability on the compositional mixing in the inner Solar System require further studies.
   
   Our simulations did not consider EJS (eccentric Jupiter and Saturn) and Grand Tack models. It has been suggested that these two models have low probability in producing distinct isotopic compositions of Earth and Mars \citep{Mah2021, Mah2022b, Woo2022}. Alternatively, the pebble accretion model could reproduce the small mass and rapid accretion of Mars \citep{Levison2015, Johansen2021}. The contribution of pebble accretion to the formation of terrestrial planets and asteroids is not considered in our simulations. The compositional effects of the amount and the timescale of pebble accretion for terrestrial planets have been investigated very recently \citep{Johansen2021, Mah2022a}. Nevertheless, how pebble accretion affects the spatial distribution of isotopes is still under debate. It has been suggested that a large mass contribution of pebbles in terrestrial planets could replace and homogenize isotopic distribution in the inner disk, with the result that the observed isotopic dichotomy could not be preserved \citep{Mah2022a}.
   
   \subsection{Implications for isotopic evolution in the protoplanetary disk}
   The spatial gradient of Cr isotopic anomaly and the absence of a $\Delta^{17}$O gradient verified by our simulations provide insights into isotopic evolution in the protoplanetary disk. The Cr isotopic anomaly may reflect incomplete mixing of different nucleosynthetic inputs in the solar nebula \citep{Dauphas2016}. In this scenario, the outer disk is dominated by the early infalling materials, and the inner disk is dominated by the late infalling materials \citep{Jacquet2019, Nanne2019}. Assuming a continuous change of the isotopic compositions for Cr, Ti, and Ca in  infalling materials, a spatial isotopic gradient in the disk can be reproduced \citep{Jacquet2019}. Alternatively, the thermal processing of molecular cloud material after complete isotope homogenization could have resulted in a preferential loss of thermally unstable presolar carriers of isotopic anomalies and in unmixing of isotopic components \citep{Trinquier2009, VanKooten2016}. The different extent of thermal processing could have resulted in the NC-CC dichotomy \citep{Trinquier2009, Schiller2018, Johansen2021} and the separation of NC and CC reservoirs could have been maintained for several million years by the viscous expansion of the protoplanetary disk \citep{Liu2022}. In this scenario, the progressive thermal processing of presolar materials led to the isotopic correlations in the NC meteorites \citep{VanKooten2016}. Various degrees of thermal processing in the solar nebula may have resulted in the correlation between the abundances of different presolar components and the chemical compositions among different chondritic groups \citep{Huss2003}. The gradient of isotopic anomalies suggested by our simulations could reflect progressive thermal processing in different regions with a temperature gradient in the protoplanetary disk. The regions closer to the Sun could have experienced a higher extent of thermal processing, evaporating a larger amount of presolar carriers that are thermally unstable and enriched in $^{54}$Cr. The preservation of $^{54}$Cr-enriched presolar carriers could have resulted in the higher $\varepsilon^{54}$Cr the further away from the Sun.
   
   The $\Delta^{17}$O variation in extraterrestrial samples has often been attributed to mass-independent fractionation processes such as CO self-shielding \citep{Yurimoto2004} and symmetry-dependent reactions \citep{Thiemens2021}, in addition to nucleosynthetic inheritance \citep{Clayton1973}. In the predissociation and self-shielding model \citep{Yurimoto2004}, two reservoirs could have been produced by CO self-shielding, with water ice depleted in $^{16}$O (high $\Delta^{17}$O) and the remaining CO enriched in $^{16}$O (low $\Delta^{17}$O) relative to the primordial oxygen isotopic compositions of the solar nebula. The $\Delta^{17}$O of the inner planetesimal disk may have been altered by water that formed in the interstellar medium \citep{Yurimoto2004, Alexander2017} or in the outer Solar System \citep{Lyons2005} with high $\Delta^{17}$O. The D/H ratio of water in carbonaceous chondrites is consistent with the interstellar origin of water, but inconsistent with water influx from the outer Solar System \citep{Alexander2012, Alexander2017}. Therefore, the $\Delta^{17}$O distribution in the inner disk could have been dependent on water ice transport, water evaporation, and water-rock interaction \citep{Yurimoto2004}, some of which may not be related to heliocentric distance. The decoupling of the $\Delta^{17}$O and the nucleosynthetic isotopic anomalies of Earth, Mars, and Vesta (Fig.~\ref{fig:1}) can be explained by the incorporation of water. The correlation between $\Delta^{17}$O and nucleosynthetic isotopic anomalies in NC parent bodies is weaker than that in carbonaceous chondrites \citep[e.g.,][]{Warren2011, Shuai2022}. The high $\Delta^{17}$O in accreted water could have been superimposed on the nucleosynthetic $\Delta^{17}$O variation in the NC parent bodies, altering the original spatial distribution of $\Delta^{17}$O and the correlation between $\Delta^{17}$O and nucleosynthetic isotopic anomalies. In summary, the absence of a spatial $\Delta^{17}$O gradient reflects the complex origin of $\Delta^{17}$O variation in the NC meteorites, which might result from both nucleosynthetic inheritance and mass-independent fractionation.
   
\section{Conclusion}
   We performed N-body simulations on the formation of the terrestrial planets and main-belt asteroids in the inner protoplanetary disk for the classical model and for the depleted disk model with two different sets of parameters (DP-1 and DP-2). The gravitational interactions between planetesimals were included in the simulations, making it possible to simulate the accretion process from a disk of planetesimals and the formation of asteroids. The distributions of modeled $a_{\rm weight}$ values are compared with the observed $\varepsilon^{54}$Cr and $\Delta^{17}$O of Earth, Mars, and the NC meteorite parent bodies. In all three models, Earth and Mars analogs have smaller $a_{\rm weight}$ values than most asteroid analogs, consistent with the observed distribution of $\varepsilon^{54}$Cr among Earth, Mars, and the NC meteorite parent bodies, but inconsistent with the observed distribution of $\Delta^{17}$O. The simulation results were further calibrated by statistical comparison between the isotopic compositions of the NC meteorite parent bodies and the modeled $a_{\rm weight}$ of asteroid analogs, assuming spatial gradients of $\varepsilon^{54}$Cr or $\Delta^{17}$O in the protoplanetary disk. The assumed linear gradients were calibrated using the width and the position of isotopic data distributions of NC meteorite parent bodies and the simulated $a_{\rm weight}$ values of asteroid analogs. Using the calibrated isotopic gradients, the Cr isotopic compositions of Earth and Mars were calculated. The calculated $\varepsilon^{54}$Cr values of Earth and Mars are consistent with the published $\varepsilon^{54}$Cr data. These results suggest that a spatial $\varepsilon^{54}$Cr gradient approximated as a linear function could have existed in the protoplanetary disk. The spatial gradient of $\varepsilon^{54}$Cr can be extended to other isotopic anomalies such as $\varepsilon^{48}$Ca, $\varepsilon^{50}$Ti, and $\varepsilon^{94}$Mo. However, the published $\Delta^{17}$O data could not be reproduced by our simulations. Values for $\Delta^{17}$O may not vary systematically with heliocentric distance in the inner disk. The spatial gradient of isotopic anomalies reflects inheritance of the isotopic composition change from the early infall to the late infall, or the progressive thermal processing of presolar materials. On the other hand, the absence of a $\Delta^{17}$O gradient reflects the complex evolution of $\Delta^{17}$O in the protoplanetary disk. In addition to nucleosynthetic variations, the $\Delta^{17}$O variations associated with mass-independent fractionation processes could have altered the spatial distribution of $\Delta^{17}$O in the inner protoplanetary disk.
   
\begin{acknowledgements}
      This work was supported by the B-type Strategic Priority Program of the Chinese Academy of Sciences (XDB41000000), National Natural Science Foundation of China (NSFC) grant (42125303), and China National Space Administration (CNSA) grant (D020205). We thank the anonymous reviewer for detailed and constructive comments and suggestions which improved the quality of the paper. We are grateful to Simon Grimm for help on N-body simulations using GENGA. We acknowledge the computational support from High-Performance Computing Center (HPCC) of Nanjing University and High-Performance Computing Center of Collaborative Innovation Center of Advanced Microstructures.
\end{acknowledgements}

\bibliographystyle{aa} 
\bibliography{reference.bib}

\begin{appendix} 
\onecolumn
\section{Isotopic data from literature}
\begin{table}[htp]
    \caption{Mass-independent isotopic variations among the noncarbonaceous meteorite parent bodies.}
    \label{table:A1}
    \centering 
    \begin{tabular}{l l c c l c c l}
    \hline\hline
     & & $\Delta^{17}{\rm O}$ & 2sd & References & $\varepsilon^{54}{\rm Cr}$ & 2se & References \\
    \hline\
        NC chondrites & H & 0.73 & 0.18 & 1 & $-$0.36 & 0.10 & 17, 18, 19 \\
         & L & 1.07 & 0.18 & 1 & $-$0.36 & 0.08 & 17, 18, 20\\
         & LL & 1.26 & 0.24 & 1 & $-$0.43 & 0.05 & 17, 18\\
         & EH & $-$0.02 & 0.34 & 1 & 0.04 & 0.05 & 17, 18, 21, 22, 23\\
         & EL & 0.00 & 0.13 & 1 & 0.04 & 0.05 & 17, 18, 21\\
         & Rumuruti chondrites & 2.72 & 0.62 & 2 & $-$0.06 & 0.02 & 24, 25\\
         \hline
        Planetary bodies & Mars & 0.32 & 0.03 & 1 & $-$0.16 & 0.03 & 17, 18, 26\\
             & Earth & 0.007 & 0.012 & 3 & 0.08 & 0.02 &
            7, 17, 18, 19, 21, 25,\\
            &&&&&&&27, 28, 29, 30, 31\\
         & Moon & 0.010 & 0.012 & 3 & 0.13 & 0.09 & 18, 21\\
         \hline
        \multirow{2}{*}{\shortstack[l]{Parent bodies of\\ NC primitive  achondrites}}
            & Acapulcoites-Lodranites & $-$1.12 & 0.36 & 1 & $-$0.58 & 0.09 & 11, 20, 24, 32\\
         & Brachinites & $-$0.25 & 0.18 & 1 & $-$0.46 & 0.16 & 33\\
         & Winonaites & $-$0.51 & 0.08 & 1 & $-$0.52 & 0.06 & 20, 32\\
         & Ureilites & $-$0.96 & 1.00 & 1 & $-$0.91 & 0.04 & 24, 30, 32, 34, 35, 36, \\
            &&&&&&&37
             \\
             \hline
        \multirow{3}{*}{\shortstack[l]{Parent bodies of\\ NC magmatic achondrites\\ and stony-iron meteorites}} &
         Vesta (HED meteorites
            & $-$0.24 & 0.02 & 1 & $-$0.69 & 0.06 & 17, 18\\
         & -mesosiderites)\\
         & Angrites & $-$0.072 & 0.014 & 1 & $-$0.43 & 0.05 & 17, 19, 24, 28, 38\\
         & Aubrites (main-group) & 0.012 & 0.014 & 1 & 0.07 & 0.06 & 17, 31\\
         & Shallowater & $-$0.02 & 0.04 & 4 & $-$0.12 & 0.04 & 31\\
         & NWA 5363 / 5400 & 0.00 & 0.07 & 5, 6, 7 & $-$0.32 & 0.06 & 7, 39\\
         & NWA 8054 & $-$1.76 & 0.04 & 8 & $-$0.44 & 0.08 & 33\\
         & NWA 7325 & $-$0.76 & 0.15 & 9, 10, 11 & $-$0.57 & 0.06 & 11, 40\\
         & Ibitira & $-$0.065 & 0.014 & 12, 13 & $-$0.40 & 0.08 & 41\\
         & Asuka 881394, Bunburra
            & $-$0.13 & 0.05 & 1, 13 & $-$0.40 & 0.06 & 42, 43, 44\\
         & Rockhole, Emmaville,\\
         & Dho 007, EET 92023 \\
         & Pasamonte, PCA 91007 & $-$0.204 & 0.011 & 1, 13 & $-$0.28 & 0.12 & 41\\
         & NWA 1240 & $-$0.271 & 0.014 & 13 & $-$0.63 & 0.10 & 41\\
         & NWA 11042 / 4284 & 1.02 & 0.04 & 1, 14 & $-$0.29 & 0.08 & 14\\
         & LEW 88763 & $-$1.19 & 0.02 & 15 &  &  & \\
         & NWA 2353 / 2635 & 0.54 & 0.27 & 5 &  &  & \\
         & NWA 2993 & $-$0.19 & 0.03 & 5 &  &  & \\
         & NWA 5297 / 6698 & 1.21 & 0.08 & 5 &  &  & \\
         & NWA 5517 & $-$0.90 & 0.02 & 5 &  &  & \\
         & NWA 11916 & $-$0.45 & 0.03 & 5 &  &  & \\
         & Pallasites (Krasnojarsk) & $-$0.18 & 0.04 & 16 & $-$0.72 & 0.10 & 17\\
         & Pallasites (Brenham) & $-$0.19 & 0.05 & 16 & $-$0.06 & 0.11 & 18\\
    \hline
    \end{tabular}
    \tablebib{
    (1) \citet{Greenwood2017}; (2) \citet{Bischoff2011}; (3) \citet{Greenwood2018}; (4) \citet{Newton2000};
    (5) \citeauthor{MetBullDatabase}; (6) \citet{Day2012}; (7) \citet{Burkhardt2017}; (8) \citet{Irving2014}; (9) \citet{Barrat2015}; (10) \citet{Weber2016}; (11) \citet{Goodrich2017}; (12) \citet{Wiechert2004}; (13) \citet{Scott2009}; (14) \citet{Vaci2020}; (15) \citet{Greenwood2012}; (16) \citet{Greenwood2006}; (17) \citet{Trinquier2007}; (18) \citet{Qin2010}; (19) \citet{Schiller2014}; (20) \citet{Schmitz2016}; (21) \citet{Mougel2018}; (22) \citet{Zhu2020a}; (23) \citet{Zhu2021a}; (24) \citet{Larsen2011}; (25) \citet{Zhu2021b}; (26) \citet{Kruijer2020}; (27) \citet{VanKooten2016}; (28) \citet{Zhu2019}; (29) \citet{VanKooten2020}; (30) \citet{Zhu2020b}; (31) \citet{Zhu2021c}; (32) \citet{Li2018}; (33) \citet{Williams2020}; (34) \citet{Shukolyukov2006}; (35) \citet{Ueda2006}; (36) \citet{Qin2010b}; (37) \citet{Yamakawa2010}; (38) \citet{Shukolyukov2009}; (39) \citet{Sanborn2016b}; (40) \citet{Sanborn2013}; (41) \citet{Sanborn2014}; (42) \citet{Sanborn2016}; (43) \citet{Benedix2017}; (44) \citet{Wimpenny2019}.
    }
\end{table}
\newpage
\section{Mass-distance distributions at the end of simulations}
\begin{figure}[H]
\resizebox{\hsize}{!}{\includegraphics{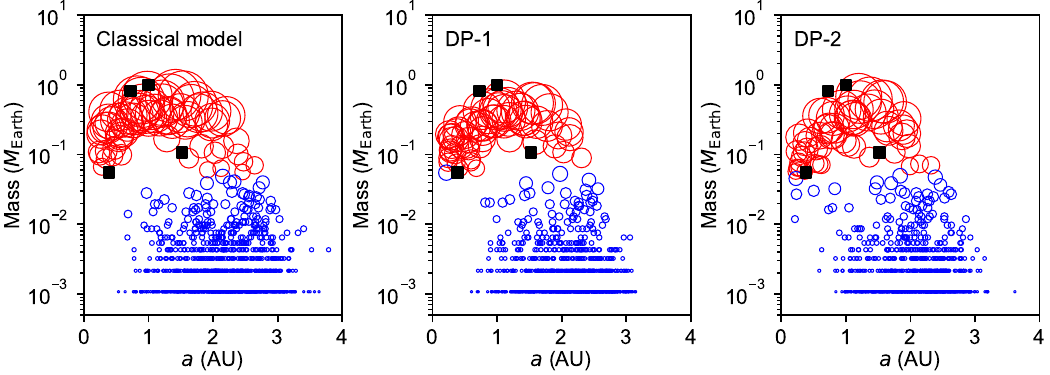}}
\caption{Mass-distance distributions of bodies at the end of all simulations. The blue circles represent the planetesimals and the red circles are the planets and embryos at the end of simulations. The size of each circle is proportional to the radius of each body but is not to scale on the distance. The black squares represent the masses and current orbits of terrestrial planets in the Solar System.
}
    \label{fig:B1}
\end{figure}

\end{appendix}
\end{document}